# LCOGT Network Observatory Operations

Andrew Pickles, Annie Hjelstrom, Todd Boroson, Ben Burleson, Patrick Conway, Jon De Vera, Mark Elphick, Brian Haworth, Wayne Rosing, Eric Saunders, Doug Thomas, Gary White, Mark Willis, Zach Walker

Las Cumbres Observatory Global Telescope, 6740 Cortona Drive, Goleta CA 93111

## ABSTRACT

We describe the operational capabilities of the Las Cumbres Observatory Global Telescope Network. We summarize our hardware and software for maintaining and monitoring network health. We focus on methodologies to utilize the automated system to monitor availability of sites, instruments and telescopes, to monitor performance, permit automatic recovery, and provide automatic error reporting. The same jTCS control system is used on telescopes of apertures 0.4m, 0.8m, 1m and 2m, and for multiple instruments on each. We describe our network operational model, including workloads, and illustrate our current tools, and operational performance indicators, including telemetry and metrics reporting from on-site reductions. The system was conceived and designed to establish effective, reliable autonomous operations, with automatic monitoring and recovery - minimizing human intervention while maintaining quality. We illustrate how far we have been able to achieve that.

**keywords:** Robotic Telescopes, Autonomous Operations, Automated Recovery, Time Domain Science

## 1. INTRODUCTION

LCOGT was conceived by our founder (WR) in the early 2000s to exploit the increasing availability of cheap computing & electronics, improving robotic control and expanding Internet bandwidth. The goal was to implement a capable, effective, reliable and sustainable global observing network; one that was designed to provide effective science without prohibitive operational support or maintenance costs. If we increase the number of telescopes and instruments by, say 30%, we want the additional science impact to exceed 30% and the cost impact to be less than 30%. Our direct operational staff currently consists of 5 people, in Australia, California and Hawaii, for 11 telescopes. Of course all of our scientific, software, engineering and IT staff, about 40 people total, contribute support, enhanced procedures, and help improve operational performance.

The original concept was for a single 2m telescope at each site, but it was quickly realized that better monitoring of multiple time-varying targets would result from having multiple smaller apertures at each site. This choice inevitably forced us to a more sophisticated scheduling system however.

The overall system should respond to high-level science observing requests as autonomously as possible. Scientific users of our system should **not** need to adjust their requests by latitude or longitude, and should not need to fill observing queues for specific sites or telescopes. Other papers describe our science operations [9149-50] and network request scheduler [9149-14]; this paper covers details of hardware, control software, and performance metrics. LCOGT is a network focused on time-varying phenomena, because that is both our scientific interest and our strength among traditional and emerging observatory trends.

The following sections summarize our hardware and software as they were designed to implement the autonomous global model. We give specific examples of our automated monitoring, fault recovery, system monitoring, and performance metrics. We describe our web-based scientific support for both internal and external users, and briefly summarize operational details like workloads, maintenance and spares control.

## 2. HARDWARE

### 2.1 Telescopes

LCOGT operates two 2-m telescopes, at Haleakala Observatory (HO) in Hawaii and at Siding Spring Observatory (SSO) in Australia. We also operate nine 1-m telescopes, three at CTIO in Chile, three at SAAO in S. Africa, two at SSO in Australia, and one at McDonald Observatory (MO) in Texas. Pending additional funding, we have additional 1-m mounts that can be deployed to these, or potentially new sites – to expand our global coverage. We are also deploying 40-cm telescopes to our existing sites. These smaller telescopes are fully capable of science observations of brighter targets, aswell as satellite and space debris monitoring. Expansion of our nodes at each site, and at new sites, is specifically included in our operational model.

The f/10 2-m telescopes have been operating for over 10-years. They are reliable alt-az telescopes, originally built by Telescope Technologies Limited (TTL) which built and deployed 5 similar telescopes worldwide. They have a cassegrain rotator to counteract field rotation, and can support up to 5 instruments, observing one at-a-time via a deployable straight-through or turn mirror. Supported instruments on our 2m telescopes currently include an optical imager, called Spectral, with 10-arcmin field of view, and a low-resolution (R~600) full-coverage spectrograph called Floyds. The latter are designed to automatically place a requested target on the slit via automatic WCS fitting and pattern matching of the field in the slit imager, then take spectra in 2 orders from 330-1050nm, eg. for SN monitoring or rapid follow-up purposes. Currently we can schedule target of opportunity observations (from request to observations) within less than 15-minutes, but this time is planned to decrease.

The f/8 1-m telescopes were designed and built by LCOGT in California. They are relatively lightweight C-ring equatorial telescopes, with Hextek mirrors. After fully assembling and testing in our warehouse, they can be deployed to already deployed domes and other infrastructure at sites, and be working on sky within 1-week of arrival, a feature of their modular design and system integration. The 1-m RC telescopes have a doublet corrector, providing over a 1-degree field in the focal plane. They support a relatively large-format optical imager called Sinistro, with 26-arcmin field of view, surrounded by 4 permanent off-axis ports (Fig. 1). The off-axis ports support an FLI (E2V) guider, space for a fast imager, and space for a fiber pickoff to a medium resolution (R~50K) spectrograph, one per site. The latter NRES spectrographs are planned to come on-line in 2015.

The 40-cm telescopes currently being deployed are also C-ring equatorial mounts, utilizing Meade RCX f/8 optics, but with all LCOGT-designed hardware, for mounts, focus and instrument control. They each have an SBIG STX6303 optical imager, with 30x20-arcmin field of view. They are mounted in pairs within a small clamshell enclosure. All our enclosures are designed to permit image quality limited by the natural site seeing, typically somewhat over 1-arcsec as our mirrors are only about 2m above ground level.

Since March 2014, all our telescopes operate under the same java-based, publish-subscribe control system, called jTCS. This provides extensive telemetry, hosted in databases on-site and in California, to help monitor, automatically control and trouble-shoot our network. LCOGT telescope details are given in Dubberley (2010)[1] and Haldeman et. al. (2010)[2]. A description of our network is given in Brown et. al. (2013)[3].

### 2.2 Instruments

Each instrument operates as a Java Agent within our Java Agent Development Environment (JADE). This provides for configured modularity for each instrument and communication between agents as necessary. There are multiple hosts, including embedded controllers for filter-wheels, motors, and power switches, and multiple communicating processes. JADE provides a yellow-pages service by which an agent can find and communicate with the other agents it may need to perform its tasks, eg. check that the dome and mirror covers are open, and the telescope is tracking and guiding. A new instrument must be fully configured, but can be fairly easily attached to a working telescope system. A new site or new telescope at a site can relatively quickly inherit its necessary configurations, and function within our network. We currently support four different types of telescopes and 10 different types of instrument agents, although some of these are superseding old ones.

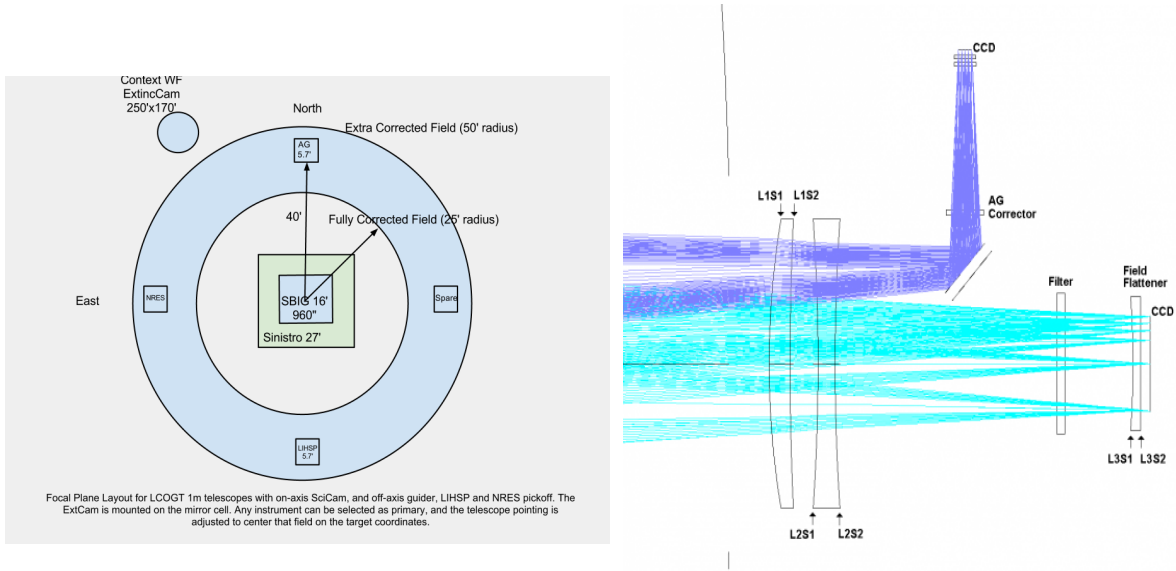

Figure 1 (Left) shows a schematic instrument layout of the LCOGT 1m focal plane; (Right) shows optical path from the mirrors, through the doublet correcter to the main science field, with provision for 4 pickoff mirrors (one is shown) for permanent off-axis instrumentation.

Our 2m and 40-cm telescopes have only on-axis instrumentation (including fold mirrors) whereas our 1m telescopes have a 40-arcmin diameter on-axis field, with up to 4 smaller (6-arcmin diameter) off-axis fields. Configuring these focal planes is handled through our jTCS system with TPK (Terrett[4]) so that selecting an instrument causes the telescope pointing to be configured to that part of the focal plane.

## 2.3 Site Infrastructure & Expansions

Each 2m site has an LCOGT staff member to provide support, maintenance and upgrades. Other sites were selected at existing observatories with good existing infrastructure, and a willingness to host our relatively small footprint. We contract with local site services to provide basic maintenance, including $CO_2$ cleaning of mirrors, routine checks and support, including trouble-shooting on demand. Considerable thought went into designing each site for ease of build, maintenance and future development. There are typically three 6-m Ash domes for three 1-m telescopes, three rectangular concrete pads to provide room for small clamshell enclosures (2mx4m, called Aqawans), each capable of supporting two 40-cm mounts, a single 20-ft air-conditioned Site Services container that provides air-conditioned computer services, safety and weather monitoring, UPS backup, and distributes power and internet for our whole site. There is also a larger container for storage of tools and parts. A weather mast provides standard weather feeds to decide when it is safe to open enclosures. Adding new telescopes or instruments to a site requires some updates to that site's software configuration, but each site is relatively easily scalable. A new site requires installation of concrete pads, conduits, domes, enclosures and Site Services container, and connection to the local site power and internet (see Pickles et. al.[5]). We have a system of Fortress keyed interlocks so our system knows when people are on-site, in the site-services container, in any enclosure, or within the Aqawan fence. All normal operations proceed when people are present, except slewing speeds are reduced. Telescopes can be locally switched to Manual or Disable during site work.

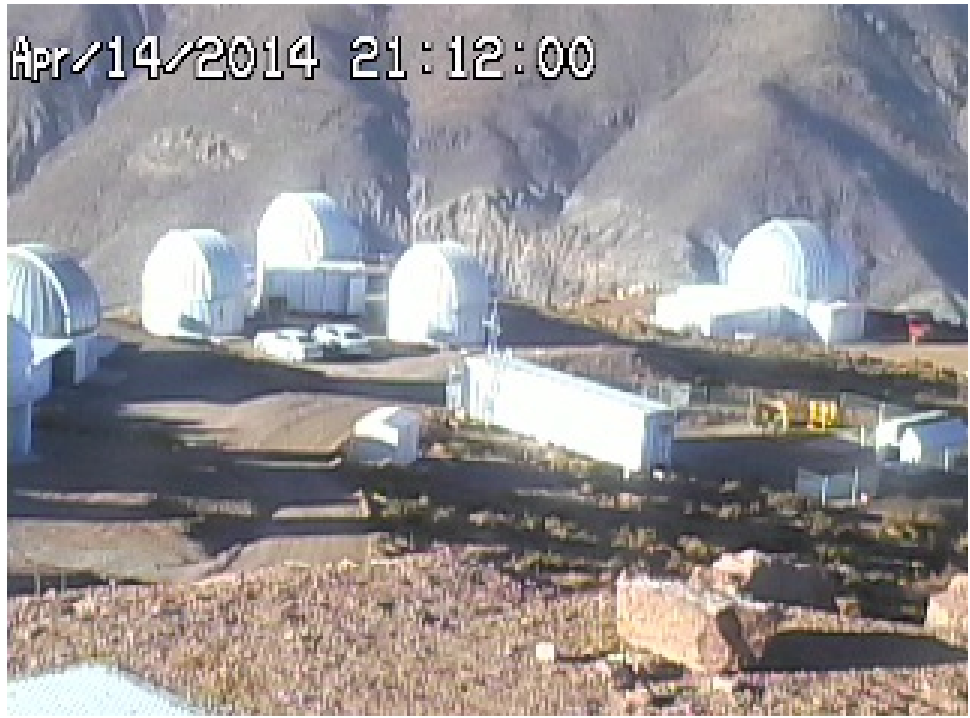

Figure 2. shows a recent image from our CTIO site, with the SMARTS 1.3m at the middle-left edge, three LCOGT 1-m domes with our small site services container (and 2 parked cars) between them, a longer storage container in the middle of the picture, with the weather mast attached, and pads for 40cm expansion within the fence to the right of that. One aqawan clamshell enclosure and two 40cm mounts were installed in April 2014. The doors of our domes face South in the Southern hemisphere. The Korean 1.6m KMTNet enclosure at CTIO is to the upper right (thanks to PromptCam for this image).

## 2.4 Real-Time monitoring of Sites, Telescopes & Instruments

Our jTCS control system provides extensive telemetry on all aspects of the weather at each site, together with telemetry from telescopes, instruments, filter wheels, collimation, pointing, tracking, and data flow. This high-cadence data is "harvested" to a lower cadence more appropriate for maintenance and stored in databases maintained in Santa Barbara. The harvested telemetry data provides the basis for our TelOps monitoring system, and Operation Alerts (OpAlerts) controlled via Nagios monitoring, that are directed to appropriate staff as warnings, and to enable human interaction and correction when necessary.

Much of our telemetry information is presented on a public web interface at http://telops.lcogt.net Yellow background on these views indicates a problem, which is described in a pop-up window when the cursor is hovered over that part. If instrument agents cannot resolve problems, then OpAlerts requiring human intervention are emailed to relevant people.

For each site view within TelOps, there is a graphing tool that shows for each telescope: the median FWHM, any mis-pointing, median ellipticity, Sun and sky conditions, including atmospheric transparency derived from Boltwood Sky-Ambient temperature sensors and expressed as a percent, Air temperature and humidity, and wind speed.

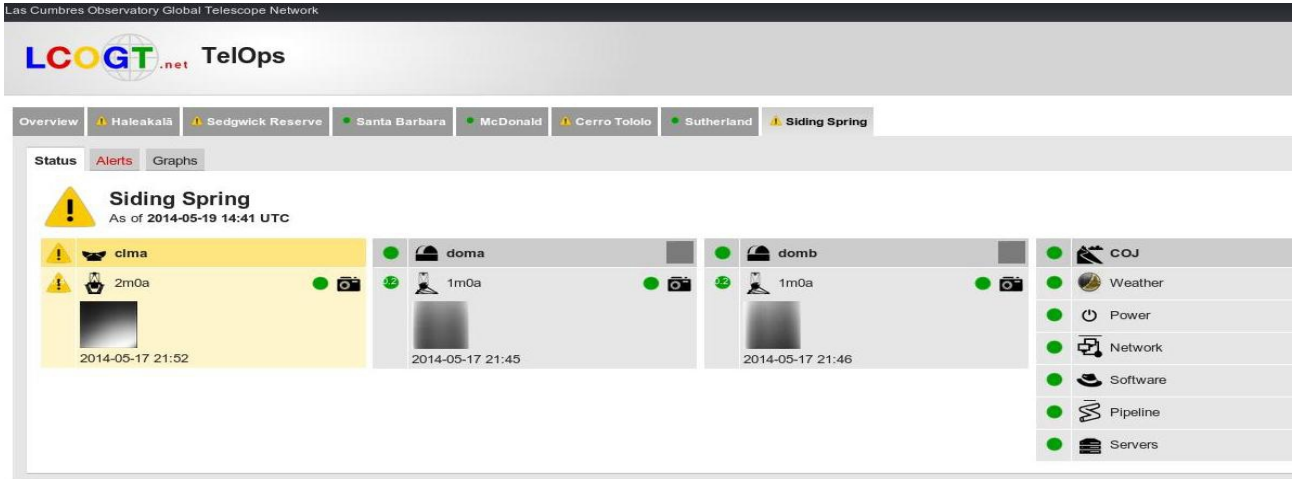

Figure 3. is a screenshot of our telescope operations website, showing the current status of our Siding Spring (COJ) site, enclosures, and telescopes, including instruments and thumbnails of most recent (raw) images.

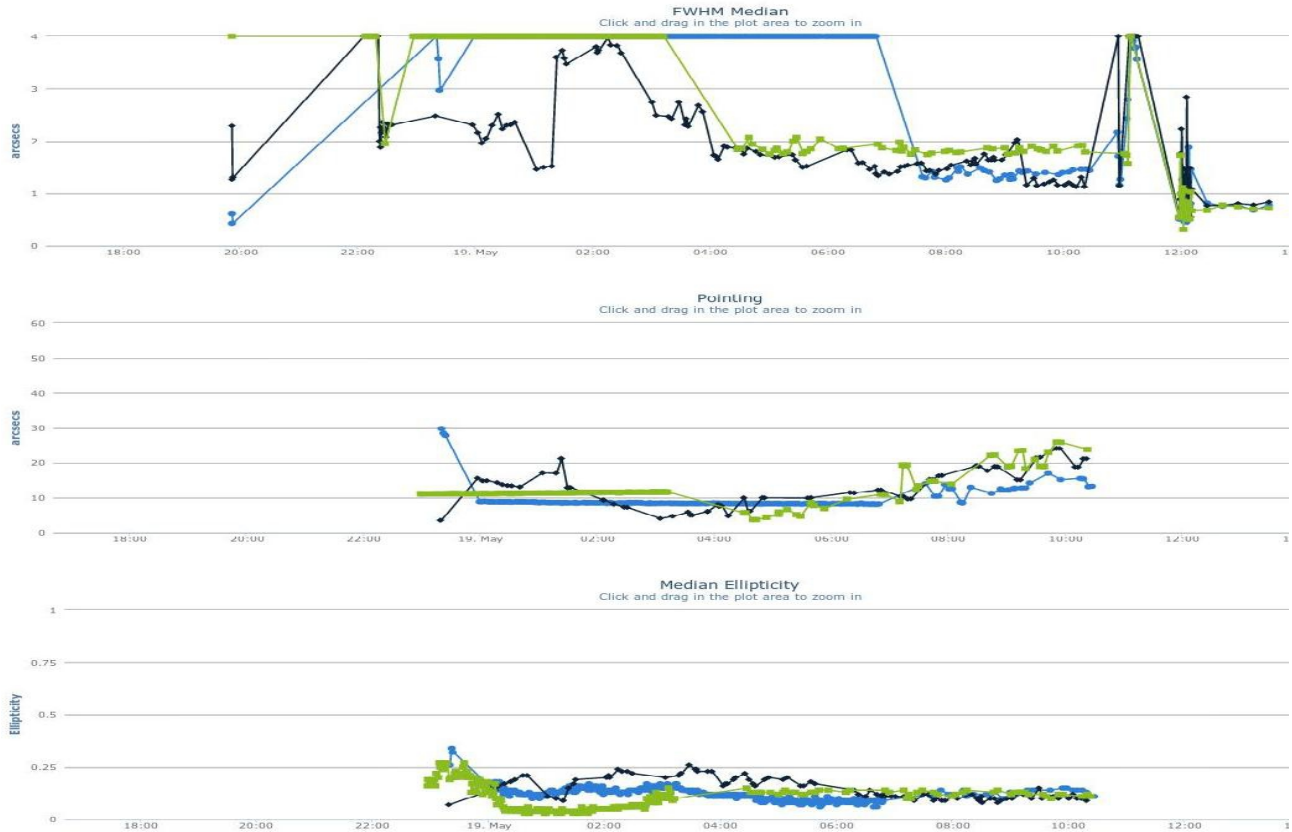

Figure 4. The FWHM graph (top) shows image results, from Sextractor, for three 1-m telescopes at CTIO during a typical night. The clipped values at 4-arcsec FWHM are when scheduled observations deliberately request a telescope defocus, typically 1-3mm in the focal plane, to avoid saturation of target or calibrating stars (eg. for planet transits of relatively bright stars). Our guiders are on independent focus stages, so they still guide in focus while the science camera is defocused. The pointing graph (middle) summarizes the typical pointing errors in arcsec. The bottom graph shows measured ellipticity results for each science image as they are taken.

At each site we quickly analyze each image as soon as it is taken at site with SExtractor[6] to measure median star FWHM and ellipticity for each science frame. We also fit them with Astrometry.Net to determine how far from the target coordinates our pointing is. This "flash" data analysis (separate from our pipelined data products seen by users) provides useful immediate feedback on telescope performance, and can trigger corrective action if necessary. The site telemetry and "flash" reduction provide both daily and long-term metrics that we use to monitor and evaluate our system. In the early deployment phase of our network, much work went into correcting problems as they manifested themselves. We are still enhancing our operations, but now we spend more time thinking about and developing long-term improvements.

The graph page (per site) also displays weather information (Fig. 5). The top panel shows daytime solar irradiance (filled light blue) with solar power scale to the left, the nighttime sky brightness measured by a Unihedron Sky Quality Meter (yellow fill - mag/arcsec$^2$ scale to the right), the estimated sky transparency from our Boltwood monitors (10um sky-ambient temperature - blue trace, 0 to 100 %, scale to the right). The effects of clouds (low transparency) on the measured solar irradiance is clearly seen in the blue-filled solar illumination trace to the right. The green trace shows the Lunar Zenith distance in degrees, on this night for a 64% illuminated Moon. The effect of the rising & setting Moon on the sky brightness can be clearly seen. The middle panel (Fig 5) shows air temperature (green), dew-point (red) and Boltwood (yellow, all left scale) and humidity (purple, right scale). All these values are clearly identified by hovering a mouse over them in the live display; an area can be expanded by clicking to select a box.. The lowest panel shows wind speed (in m/s), with the gust speed in darker blue. Our wind speed threshold for closing enclosures is 15 m/s (54 kph).

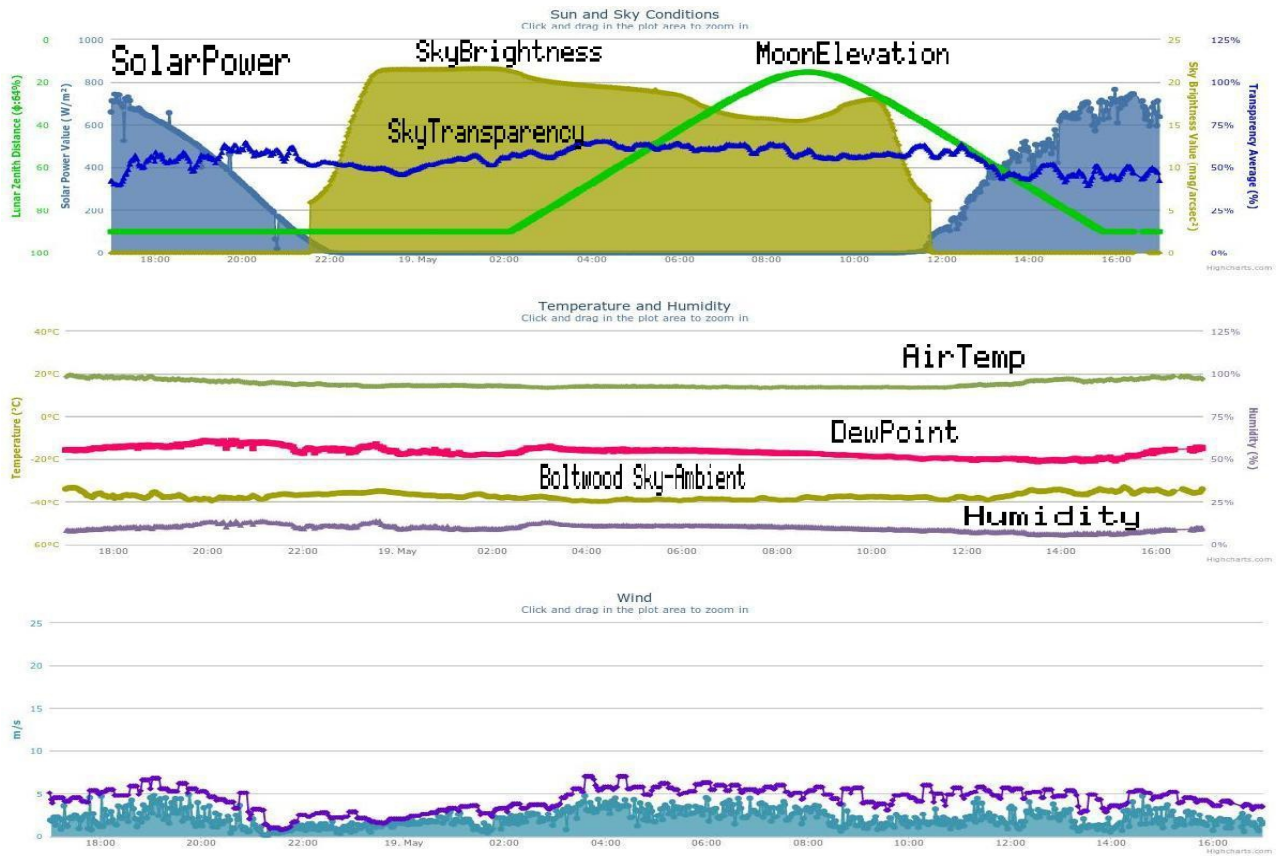

Figure 5. Site weather graphs: Sun and Sky conditions (top), Temperature & Humidity (middle), Wind (bottom). See text for details.

While this website can be (and is) perused by people associated with LCOGT operations, the real strength lies in the automated Operations Alerts. These include severe weather warnings, latency issues caused by internet problems, and telescope or instrument issues that cannot be auto-recovered by the Java Agents. The only Alert current for this site at this time was one for high telemetry latency. We have had situations where high latency (minutes to hours) prevents us from actively monitoring a site, but sites can continue normal operations: opening, observing from its local copy of the schedule (for up to 72 hours in the event of complete network outage), collecting data, and automatically closing for weather or daytime.

## 3. PERFORMANCE METRICS

### 3.1 Site Performance

Fig. 6 shows a high-level metric, useful for both operations and reporting to partners, TACs and SACs. It shows hours that our sites and telescopes spend down for weather, engineering (both maintenance and fault fixing), the amount of clear time that is scheduled for observations, and the amount of open-shutter time.

**LCOGT.net TelOps Metrics**

From: 2014-05-04  To: 2014-05-19  Go

|  | 1-meter Network | | | | | | | | | 2-meter Network | |
|---|---|---|---|---|---|---|---|---|---|---|---|
|  | ELP | COJ | | LSC | | | CPT | | | | |
|  | DOMA | DOMA | DOMB | DOMA | DOMB | DOMC | DOMA | DOMB | DOMC | FTN | FTS |
| Total Hours | 360 | 360 | 360 | 360 | 360 | 360 | 360 | 360 | 360 | 360 | 360 |
| Nighttime Hours | 126.47 | 172.56 | 172.56 | 171.97 | 171.97 | 171.97 | 173.15 | 173.15 | 173.15 | 137.73 | 172.56 |
| Weathered-out Hours | 22.9 | 110.26 | 110.26 | 48.19 | 48.19 | 48.19 | 106.09 | 106.09 | 106.09 | 37.19 | 110.26 |
| Maintenance Hours | 0 | 0.03 | 0.03 | 0.15 | 0 | 0 | 0 | 0 | 0 | 26.65 | 0.01 |
| Calibration Hours | 32.96 | 24.2 | 24.33 | 38.99 | 41.09 | 42 | 23.08 | 22.63 | 21.58 | 42.62 | 30.74 |
| Usable Science Hours | 103.57 | 62.27 | 62.27 | 123.63 | 123.78 | 123.78 | 67.06 | 67.06 | 67.06 | 73.89 | 62.29 |
| -- FTP | 2.16 | 0.83 | 3.08 | 0 | 0 | 0.3 | 0 | 0 | 0 | 3.03 | 3.16 |
| -- HAWOL | 0 | 0 | 0 | 0 | 0.41 | 0.41 | 0.02 | 0.02 | 0.02 | 0.37 | 0 |
| -- KEY2014A-002 | 15.9 | 0.4 | 1.07 | 0 | 0 | 0 | 2.25 | 1.22 | 0.18 | 2.48 | 0.7 |
| -- KEY2014A-003 | 21.35 | 1.58 | 0.67 | 0.82 | 0.83 | 0.82 | 2.37 | 1.61 | 1.91 | 7.18 | 2.67 |
| -- KEY2014A-004 | 0 | 8.23 | 8.78 | 23.69 | 26.84 | 14.95 | 13.67 | 8.01 | 2.1 | 0 | 0 |
| -- LCOGT | 14.13 | 19.46 | 16.78 | 26.09 | 24.64 | 39.61 | 15.17 | 14.7 | 29.32 | 8.34 | 3.56 |
| -- LCOeng | 0 | 0 | 0 | 0.6 | 0 | 0 | 0 | 0 | 0 | 0 | 0 |
| -- SAAO | 0.27 | 0.12 | 0.05 | 0 | 0 | 0 | 0.42 | 0.28 | 0.62 | 0 | 0 |
| -- SUPA | 0 | 0.03 | 0.08 | 0 | 0 | 0.39 | 0.06 | 0 | 0.07 | 0 | 0 |
| Total Science Hours | 53.81 | 30.65 | 30.51 | 51.2 | 52.72 | 56.48 | 33.96 | 25.84 | 34.22 | 21.4 | 10.09 |
| Idle Hours | 49.76 | 31.62 | 31.76 | 72.43 | 71.06 | 67.3 | 33.1 | 41.22 | 32.84 | 52.49 | 52.2 |
| Open Shutter Hours | 43.88 | 22.49 | 19.61 | 24.74 | 24.21 | 20.13 | 24.06 | 18.01 | 21.84 | 19.07 | 11.62 |

Download as CSV

**Legend**

**Total Hours:** Number of hours in the period.
**Nighttime Hours:** Nautical night hours, by site, over the period.
**Weathered-out hours:** The sum of all nighttime intervals during which we were closed for weather over the period.
**Maintenance hours:** The sum of all nighttime intervals during which we were not closed for weather but the telescope sequencer was disabled, over the period.
**Calibration Hours:** The sum of all intervals for which we had a successful or failed calibration over the period (usually occurs during daytime).
**Usable Science Hours:** Nighttime hours - Weathered-out hours - Maintenance Hours.
**Partner Hours:** The sum of all attempted observations for that partner on a given telescope, over the period.
**Idle Hours:** Usable Science Hours - sum(Partner Hours).
**Open Shutter Hours:** Based on the framedb, this is the sum of all exposure open shutter hours for the science instrument, over the period.

Figure 6. summarizes hours lost to daytime, weather, maintenance/engineering for a specific period.

Hours lost to weather (Fig. 6) are usually beyond control, although the opening criteria in the presence of clouds can be changed. Currently we close when the transparency is measured less than 25%, and re-open when it is >30%. This is acceptable for some programs, less so for others. We do not provide automatic exposure timing to adjust for clouds.

Calibration hours here refers to taking of bias and dark frames during daytime in the closed and dark enclosures (our instruments are also well darkened) , and twilight flat-field calibrations, typically taken 105 degrees away from the Sun, with the Sun zenith angle between 88 and 98 degrees. Science time here refers to all hours with the Sun below 102 degrees (nautical twilight) as many of our brighter targets can be observed before full astronomical twilight.

We are of course working to reduce the technical outages, particularly at our 2m sites where we have had ongoing telescope start-up issues, some of which is related to the older hardware on these telescopes. We are also working to improve the scheduled fraction within clear night-time (see [9149-14]) and our open-shutter fraction on sky. The latter is a combination of reducing our instrument read-times, and combining more functionality in parallel between science frames.

Our biggest effort to date has been on ensuring the quality (rather than quantity) of observed data. Our operational model charges users for all data taken on their project(s), including if data has to be re-taken because it was a) not completed or b) didn't pass quality control because of mis-pointing, poor image quality, or low S/N. The latter might be due to inadequate exposure calculations (performed by users with the aid of a calculator) or possibly due to intervening cloud. Currently the only observing constraint checked by the scheduler is for airmass limits. In the future we will allow users to specify their observing requests with both acceptable seeing (FWHM) and transparency limits.

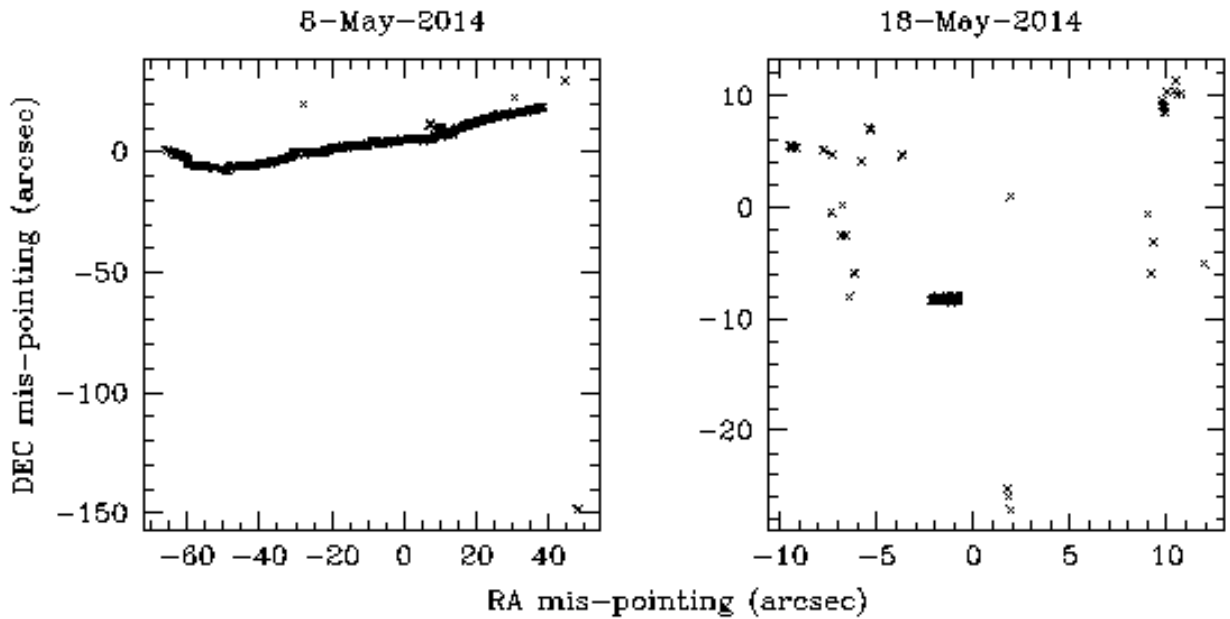

Figure 7. Comparison of pointing errors from one site in the first week of May 2014 (left), and after we fixed startup errors with our 3-axis collimators (see text for details).

### 3.2 Telescope Performance

At the start of v1 operations (1-May-2014) we knew we still had a problem with telescope mis-pointing, despite all telescopes having good Tpoint fits to about 6-8 arcsec RMS. This had 2 separate causes: 1) was that the 1m telescope 3-axis M2 collimator would occasionally fail to go properly to its default position at the start of each night (These values were measured and set during telescope deployment to provide optimum collimation) 2) is that the telescopes occasionally lose their HA/DEC index marks. The first and most common error was fixed by providing software checks at start-up: each telescope collimation is verified to be properly set, or is set again (automatically) if necessary – this also ensures optimum image quality of course. The cause of the second problem has still not been fully diagnosed, but is being dealt with by a pointing "molecule" at the start of each night, before science observing starts. The results in Fig. 7 show the improvement for one site (CTIO, LSC); others are similar.

### 3.3 User Interface

Users with accepted programs on our network can access our network through an Observation Data INterface, sometimes called ODIN, or Portal, at http://lcogt.net/observe/ and via "observatory" links on our main page at http://lcogt.net. All access to LCOGT, including an image FITSviewer, is designed to be via web-based interfaces, for ease of use, and to avoid the need for specialized software downloads by users. The user interface allows official users to submit proposals and add observation requests to accepted proposals, check the status of their observing requests and data, and to submit feedback.

All of our staff, and external users, can provide feedback highlighting an issue with network performance or data quality. Our basic procedure is to assess the feedback, generate a ticket if necessary for it (within RedMine) and assign it to an appropriate Software, Engineering or Scientific staff member to follow it through, analyze it, find and implement a solution. This way we have a good record of problems, investigations and solutions. We encourage comments about data quality, both good and bad, as these help us to improve the network.

### 3.4 System performance

The network scheduler takes requests from all users, and arranges them in a Proposed Observation Network Database (POND). The schedule changes frequently as the scheduler runs continuously, taking into account changing conditions, including site and telescope availability, and changing requests (which can be cancelled and re-submitted). The POND has an "analytics" feature that summarizes the success of each request (completed blocks vs. requested blocks) or failure modes where they occur. This provides a powerful diagnostic and operational tool, since at any time we can identify and address the most common failure modes, therefore providing the most impact of stretched software effort on network performance.

We are now using the analytics feature of our system to monitor and investigate network problems. Currently this is the method producing the most leverage on where and how network problems occur, and the most effective in reducing network errors. It is however primarily a software tool, and doesn't compare with careful data quality assessment by experienced astronomers.

**We therefore actively seek feedback from all our users, especially on data quality, via our web interface, or to sciencefeedback@lcogt.net.**

## 3.5 Analytics

Fig. 8 below shows a history of successful blocks (requests) completed on our 1-m network, for the period 1 to 21 May 2014. The black trace is the sum of all blocks, with different colored traces identifying completed blocks on each telescope. We are deliberately not over-subscribing the network yet, so this shows us completing about 70% of scheduled requests, where some fail due to changes in the weather, and some due to other problems. The network is a sophisticated system, with potential problems from the instruments, mechanisms, telescopes, telescope control, sequencer, scheduler and many other interacting parts. We tend to deal with the network **AS** a network, rather than focusing on individual telescopes or components, but methods like these do help identify both hardware and software issues, and enable us to fix problems for the longer-term. This short operating period snapshot shows that we are increasing our success rate, slowly.

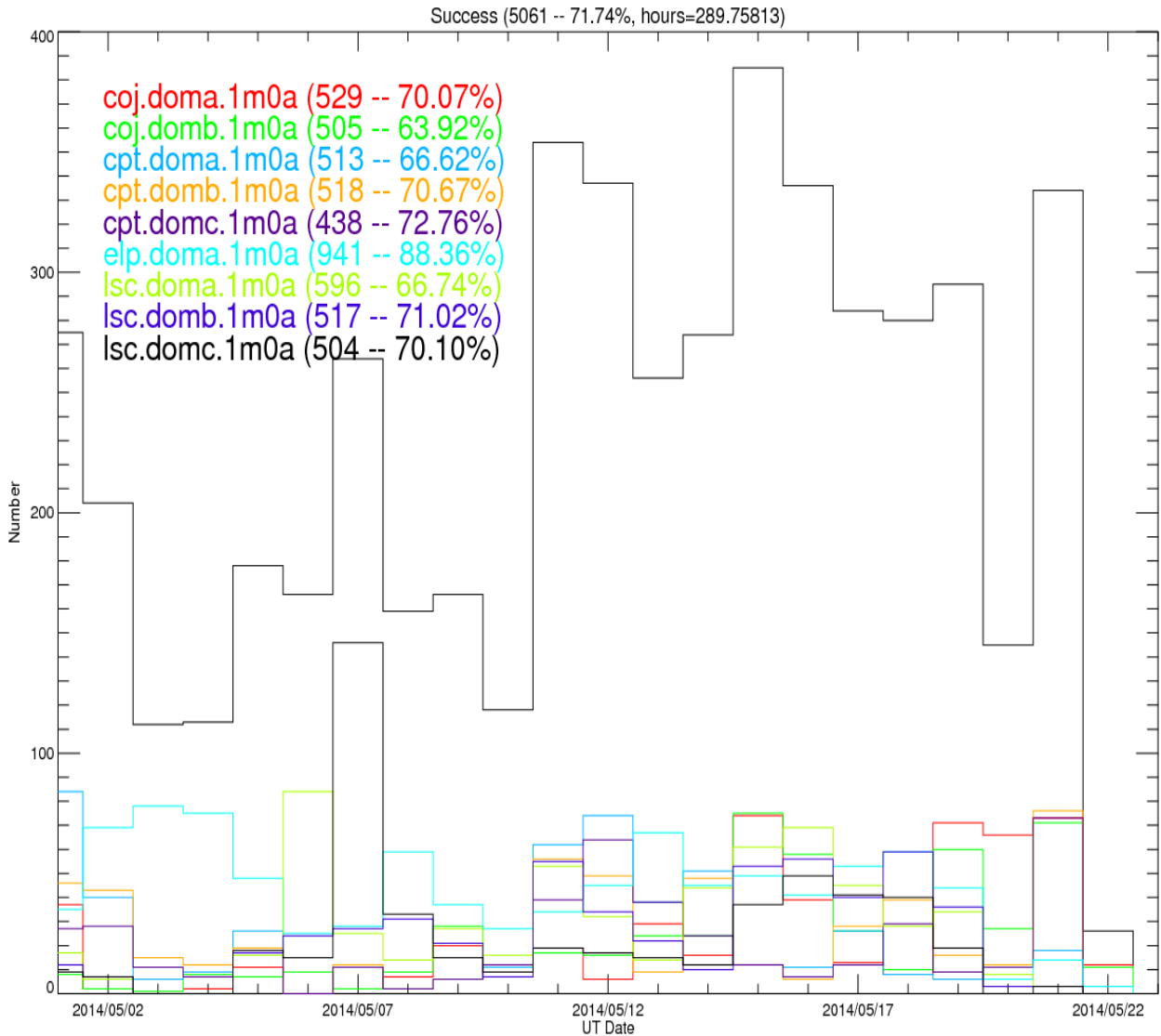

Figure 8. Histogram of successful observations on our 1m network in May 2014 (see text for details).

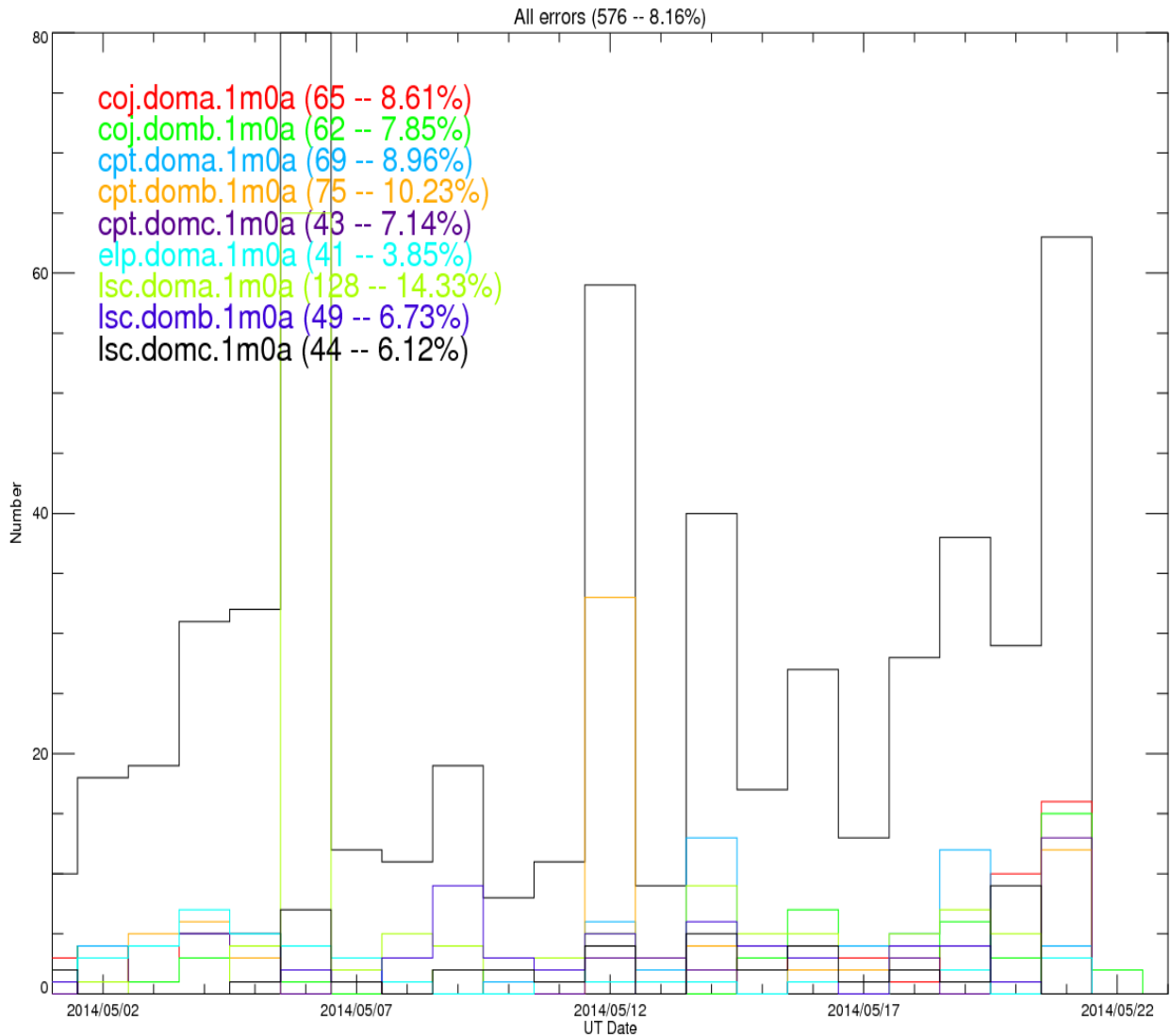

Figure 9. histogram of 1m network schedule errors during May (see text for details).

Fig. 9 above shows scheduled block errors, for the same time period, again the black trace is the total, with colored traces for errors from each telescope. The spike on 2014/05/06 comes mainly from DomA in Chile (CTIO, LSC) and was due to a specific instrument agent failure problem that was resolved the next day. The spike at 2014/05/12, mainly from DomB at CPT (Sutherland, S. Africa) is due to a problem where the science instrument was not responding appropriately within a fixed time interval. This error, while manifesting in a particular instrument, could occur again elsewhere. It was fixed with a more sophisticated code upgrade to avoid such problems. These graphs illustrate a) how we triage our problems, b) investigate the most important ones and c) by solving them in turn reduce network problems and improve performance.

## 3.6 Quality Control Steps

Examples of fairly complicated system interactions we have dealt with include a recent one with our new Sinistro imager (26 arcmin field). These are controlled by a computer, and use an image processing library to re-arrange data read from 4 readout amplifiers, compensating for the slightly different bias and gains per amplifier to achieve a single uniform 4Kx4K image. This system developed an annoying property of appearing to be working properly, but failing to write the final image. This was addressed by including more checks in the image pre-processor stages. Another issue affects our SBIG 4K imagers (15-arcmin field of view, and used in 2x2 binning mode). These imagers have a memory leak problem causing shutter imprecision, which can lead to noticeable "banding" in the images after a few hundred images are taken. Again the fix was to detect the banding in the pre-process stage, and power-cycle the CCD, if necessary during night-time observing, causing the shutter to re-home.

At each stage we have to decide whether a quick fix is in order, or whether a longer-term fix is preferred, even though that may prevent observations for some time at some telescope or site. Code changes and releases are carefully controlled and documented to maintain a full design and release history. We do however still have problems, including ones involving unforeseen consequences of some code changes.

Since we have moved from system development to Operations, we have recently introduced a more regimented system for code and configuration changes. Specific upgrades can be tested in a release branch at one site, including our prototype site in California, before being included in the general network release. Site and telescope configuration files are maintained and edited from one place. If code or configurations change at a site, that site software is restarted at local noon, to pick up all the deployed changes.

## 4. WORKLOADS

We employ one full time LCOGT person at each of our 2m sites. Both these sites now host additional telescopes, two 1-m telescopes in Australia and one ASAS mount http://www.astronomy.ohio-state.edu/~assassin/index.shtml at Haleakala. We are preparing four 0.4m telescopes in our shop in Santa Barbara, and will soon deploy two at each 2m site, housed within their large 2m clamshell enclosures. Additionally we have one full time operations staff based at Liverpool in the UK and one soon to be based in Australia. Our head of Engineering, and our Operations Scientist are based in Santa Barbara.

The head of engineering coordinates all operational activities: site deployments (including new Sinistro CCDs), on-site maintenance by visiting LCOGT staff (1 to 2 trips per year), remote troubleshooting, and coordinates with host-site observatory staff for necessary work at McDonald, SAAO and CTIO. Host site staff conduct weekly maintenance, including $CO_2$ cleaning of mirrors, and respond quickly and very effectively in our experience to requests for help with IT equipment, telescopes and instruments. Such external requests occur about 1-3 times/month/site.

Our Operations Scientist is responsible for assessing the health of the network, quality of data produced, and working with other scientific staff to analyze data and improve quality control. They are also responsible for responding to internal and external user requests for information and help with their observing requests, and responding to user feedback.

Most of our 2m spares are stored at our two 2m sites, where we have good storage space. We maintain a comprehensive suite of tools at each site, but typically only a small inventory of spares at each 1m site. We keep spare UPS batteries, embedded controller modules, and some CPU and Disk spares. We generally find it easier and cheaper to send rarely required spares to site as necessary, usually within 1 week, but we have also had success sourcing standard spares in the local countries. Our overall spare utilization has been low so far, with our 1-m telescopes starting to operate in April 2012.

## SUMMARY

We have implemented a powerful robotic, autonomous network, with good longitude coverage for time-variable phenomena. The system is sophisticated, and complicated in parts, but becoming easier to monitor and improve with a variety of software and management tools layered on the extensive telemetry and system monitoring we have in place.

Further enhancements will include heavier utilization at all sites and telescopes, with a better system of prioritized programs, including background programs, and more targets of opportunity. Our automatic acquisition system for spectra with Floyds works, but can be (and is being) improved. We will deploy more Sinistro, larger format optical imagers this year, to our 1m sites in S. Africa, Australia and Texas. We will deploy more 0.4m telescopes to several sites, and we are exploring additional funding options to expand the number of 1-m telescopes, especially to additional sites in the N. hemisphere. We are having good success with increasing science output from our telescopes. We have recently broadened our user base to include host and other partner institutions, together with some partners who have purchased time on our network. We are working hard to improve the quality of observational data for all these users.

The management tools we have in place are paying dividends in terms of actively monitoring our system, and developing more considered approaches to system improvement. We still fix problems as they arise, but as our system has stabilized, we can think more about the future.